\documentstyle[12pt,axodraw]{article}

\catcode`@=11
\@addtoreset{equation}{section}
\catcode`@=12

\def\bg#1{\mbox{\boldmath $ #1 $}}
\def\vec#1{\bg{#1}}
\def\sub#1{_{\rm #1}}

\begin{document}
 
\title {\centerline{\normalsize SINP/TNP/98-09 \hfill astro-ph/9803312} 
\bf Neutrino astrophysics and cosmology~:\\ \bf recent
developments\thanks{\sf Plenary talk given at the {\em ``B and Nu 
Workshop''}\/ held at the Mehta Research Institute, Allahabad, India,
from 4 to 8 January 1998.}}
 
\author{\bf Palash B. Pal \\
\normalsize Saha Institute of Nuclear Physics, 
Block-AF, Bidhan-Nagar, Calcutta 700064, INDIA}
 
\date{}
\maketitle

\begin{abstract}
In this talk, I have discussed some issues of recent interest and
activity in the field of neutrino astrophysics and cosmology. The
topics are: (1)~The origin of high peculiar velocities of pulsars;
(2)~Energization of the supernova shock wave; (3)~Ultra-high energy
neutrino astronomy; (4)~Possible implications of the recent
measurements of low deuterium abundance. 
\end{abstract}

\section{Introduction}
It was known, since the birth of modern astrophysics in the early part
of the 20th century, that neutrinos play an important role in various
processes that occur within a stellar core and which are responsible
for energy generation in a star. Gradually, the importance of
neutrinos were understood in stars outside the main sequence. And,
since the discovery of the microwave blackbody radiation, it was taken
for granted that there is a similar cosmic background of neutrinos,
although experimentally this background has not been detected so
far. Various constraints from neutrino properties have been deduced
from this belief, some of which are much better than the corresponding
constraints from earth-based experiments. For example, one can cite
the mass bound on stable neutrinos which are derived from the energy
density of the universe as a whole. This sets an upper bound of order
of a few tens of eV, whereas the direct measurement of the mass of
$\nu_\tau$ sets upper bounds in the range of a few tens of MeVs. If
the neutrinos are unstable, then also there exists quite severe bounds
on their lifetimes.

Unfortunately, in this talk I cannot review all of these
aspects. Rather, I will have to assume that the audience is familiar
with these concepts. The reason is that, fortunately, there has been a
lot of progress in the field of neutrino astrophysics in the last year
and a half, and quite a few of them are remarkable. I have to
concentrate on these recent developments. I cannot guarantee that I
will cover even all of the interesting recent developments. Let me say
I will cover what I know, with the restriction that I will leave out
topics such as solar neutrinos and atmospheric neutrinos, which are
covered by other speakers in this conference.

\section{Pulsar kicks and neutrinos}\label{pk}
It has been known for some time \cite{pk|obs1,pk|obs2} that pulsars
have large peculiar velocities, of the order of a few hundreds of
kilometers per second. The average value, from a sample of about a
hundred pulsars, is $450\pm90\,{\rm km\,s^{-1}}$. The reason for such
high velocities is not clearly understood.

Pulsars are rotating neutron stars which are believed to have a large
surface magnetic fields. They are born from supernova explosions. It
is not impossible that they get a kick from this explosion, provided
the supernova collapse is asymmetric. Recently, however, Kusenko and
Segr\`e have suggested a very elegant mechanism in which, even though
the matter density is spherically symmetric, the neutrino emission is
not, and this provides a clue to the understanding of the pulsar
kicks. The scenario involves some intricacies of neutrino physics, and
provides some insight into neutrino masses. In this section, we will
try to understand their idea.

Typical pulsars have masses between $1.0M_\odot$ and $1.5M_\odot$,
i.e., about $2\times 10^{33}$\,g. The momentum associated with the
proper motion of a pulsar would therefore be of order
$10^{41}\;$g\,cm/s. On the other hand, the energy carried off by
neutrinos in a supernova explosion is about $3\times10^{53}$\,erg,
which corresponds to a sum of the magnitudes of neutrino momenta of
$10^{43}\;$g\,cm/s. Thus, an asymmetry of the order of $1\%$ in the
distribution of the outgoing neutrinos would explain the kick of the
pulsars.

How could this asymmetry be generated? The key issue is the
propagation of neutrinos in a magnetic field. It is, of course,
trivially true that if neutrinos have some magnetic moment, their
motion will be affected by an external magnetic field. The more
non-trivial result, shown earlier by D'Olivo, Nieves and Pal
\cite{pk|DNP}, is that the motion of neutrinos are affected in the
presence of a background magnetic field even if they do not have any
intrinsic magnetic moment, or indeed any property that are not part of
the standard model of particle interactions. In other words, even if
the neutrinos are massless (and consequently have no intrinsic magnetic
moment), they acquire an effective magnetic moment \cite{pk|NP} due to
their weak interaction with particles in the medium. As a result, the
dispersion relation of massless neutrinos is given by~\cite{pk|DNP}
	\begin{eqnarray}
\omega = \left| \vec k - c \vec B \right| + b \,,
	\end{eqnarray}
where $c$ and $b$ depend on the distribution function of the
background electrons, whose explicit forms will be discussed
shortly. For small fields, this can be written as
	\begin{eqnarray}
\omega = K - c {\vec k \cdot \vec B \over K} + b
\,,
	\end{eqnarray}
where $K=|\vec k|$. If the free neutrinos have some mass $m\ll K$,
this relation should be modified to 
	\begin{eqnarray}
\omega = K + {m^2 \over 2K} - c {\vec k \cdot \vec B \over K} + b
\,,
	\end{eqnarray}
neglecting higher order terms in the mass. 

The extra $\vec B$-dependent term can affect resonant neutrino
conversion in the stellar core \cite{pk|reso}. To see this, we start
from the Hamiltonian governing neutrino propagation in the vacuum,
assuming a two-level system:
	\begin{eqnarray}
H = \left( \begin{array}{cc} - {\Delta m^2 \over 4K} \cos 2\theta &
{\Delta m^2 \over 4K} \sin 2\theta \\ 
{\Delta m^2 \over 4K} \sin 2\theta & {\Delta m^2 \over 4K} \cos
2\theta 
\end{array} \right) \,,
	\end{eqnarray}
where $\Delta m^2=m_2^2-m_1^2$, the mass squared difference of the two
eigenstates, and $\theta$ is the mixing angle. We have omitted a term
proportional to the unit matrix, since that is irrelevant for our
discussion.

In presence of the extra terms due to matter and magnetic field, the
Hamiltonian is modified:
	\begin{eqnarray}
\widetilde H = 
\left( \begin{array}{cc} - {\Delta m^2 \over 4K} \cos 2\theta - c_e
{ \vec {\scriptstyle k} \cdot \vec { \scriptstyle B} \over K} + b_e &
{\Delta m^2 \over 4K} \sin 2\theta \\ 
{\Delta m^2 \over 4K} \sin 2\theta & {\Delta m^2 \over 4K} \cos
2\theta 
\end{array} \right) \,.
	\end{eqnarray}
Here, once again the contribution to $b$ and $c$ from neutral current
has been omitted, since it is identical for both neutrinos. The
contributions from the charged current interactions affect only the
$\nu_e$ state. To the leading order in the Fermi constant, these are
given by~\cite{pk|DNP,pk|DNPnote}
	\begin{eqnarray}
b_e &=& \sqrt 2 G_F \left( n_e - n_{\bar e} \right) \,, \nonumber\\*
c_e &=& - 2\sqrt 2 eG_F  \int {d^3p\over (2\pi)^3 2E} \; {d\over dE}
(f_e - f_{\bar e}) \,,
	\end{eqnarray}
where $f_e$ and $f_{\bar e}$ are the Fermi distribution functions for
the electron and the positron, and $e$ is the charge of the
positron. For a degenerate electron gas at zero temperature, we can
put $n_{\bar e}=0$, and evaluation of the integral in $c_e$ yields
	\begin{eqnarray}
c_e = {eG_F \over \sqrt 2} \left( {3n_e \over \pi^4} \right)^{1/3} \,.
	\end{eqnarray}

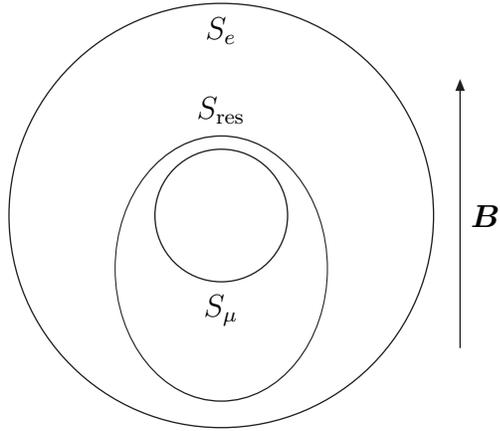
\begin{figure}
\begin{center}
\begin{picture}(110,140)(-10,-20)
\CArc(50,50)(25,0,360)
\Text(50,20)[t]{$S_\mu$}
\CArc(50,50)(80,0,360)
\Text(50,120)[]{$S_e$}
\Oval(50,30)(50,40)(0)
\Text(50,90)[]{$S\sub{res}$}
\LongArrow(140,0)(140,100)
\Text(150,50)[]{$\vec B$}
\end{picture}
\end{center}
\caption[]{\sf The relative positions of the neutrino-spheres and the
surface of resonance.}\label{nusurfaces}
\end{figure}
The condition for resonance \cite{pk|reso} is obtained by equating the
two diagonal elements of the modified Hamiltonian $\widetilde H$,
which reads 
	\begin{eqnarray}
{\Delta m^2 \over 2K} \cos 2\theta &=& b_e - c_e{\vec k \cdot \vec B
\over K} \nonumber\\*
&=& \sqrt 2 G_F n_e - {eG_F \over \sqrt 2} \left( {3n_e \over \pi^4}
\right)^{1/3} {\vec k \cdot \vec B \over K} \,. 
	\end{eqnarray}
Consider now neutrinos of a certain value of momentum. The left side
of this equation is fixed now, since $\Delta m^2$ and $\theta$ are
fundamental parameters which are not in our hands. On the right side,
the value of $n_e$ for which this equality will be satisfied will now
depend on the direction with respect to the magnetic field, because of
the quantity $\vec k \cdot \vec B$. In the direction along $\vec B$,
the resonance condition can be satisfied for a higher value of $n_e$
compared to the no-field case, i.e., at a smaller distance from the
center. In the opposite direction, since $\vec k \cdot \vec B<0$, we
need a smaller value of $n_e$, i.e., resonance will occur farther
from the center. Overall, the shape of the surface of resonance will
be ellipsoidal. A schematic section of this surface is shown in
Fig.~\ref{nusurfaces}, where this surface of resonance has been called
$S\sub{res}$.

To see how it can affect the momentum distribution of the neutrinos
coming out, let us first review the situation without any magnetic
field. In the proto-neutron star, the neutrinos near the core cannot
come out easily, because the density is so large that their mean free
path is very small. Once they reach a certain radius where the
densities are low enough, their mean free path becomes larger than the
radius of the proto-neutron star and they can escape. The surface at
this radius is called the neutrino-sphere.  Since the cross section of
$\nu_e$ with matter is higher than that of $\nu_\mu$ owing to charged
current interactions, the neutrino-sphere for the $\nu_e$'s is at a
smaller density, i.e., larger radial distance, than that for the
$\nu_\mu$. These two neutrino-spheres are schematically shown in
Fig.~\ref{nusurfaces} with the symbols $S_e$ and $S_\mu$.

Let us now see, after Kusenko and Segr\`e, how this picture might
change in presence of magnetic fields. We have discussed the surface
of resonance, $S\sub{res}$. Suppose now this surface lies in between
the $\nu_e$ neutrino-sphere and the $\nu_\mu$ neutrino-sphere, as has
indeed been shown in Fig.~\ref{nusurfaces}. The $\nu_\mu$'s produced
in the core would escape before they reach this surface. The
$\nu_e$'s, however, can convert resonantly to $\nu_\mu$'s at the
surface of resonance. Since at this point, they are outside the
$\nu_\mu$ neutrino-sphere, they will escape the star once this
conversion takes place.

Now comes the crucial point. In directions where the resonance surface
is close to the center, the neutrinos come out with larger average
momentum, since the temperature there is larger. In opposite directions
where the resonance surface is far from the center, the neutrinos have
smaller average momentum. This creates the momentum imbalance, and the
pulsar gets a kick. Analysis of the situation shows that in order to
get a fractional imbalance of the order of $1\%$, one needs magnetic
fields of the order of $3\times 10^{14}$\,G, which does not look at
all improbable inside a proto neutron star, for which surface fields
are of order $10^{12}$---$10^{13}$\,G.

One condition for this picture to work is that, as stated earlier, the
surface of resonance has to lie between the two neutrino-spheres. For
small values of the mixing angle $\theta$, this implies that
	\begin{eqnarray}
\Delta m^2 \sim 10^4 \; {\rm eV}^2 \,.
	\end{eqnarray}

Of course, in the entire discussion, it has to be understood that it
does not matter whether the resonant conversion takes the $\nu_e$ to
$\nu_\mu$ or to $\nu_\tau$. But in any case, the value of $\Delta m^2$
indicated above is in conflict with cosmological bounds on stable
neutrino masses, and also is not suggested by any other indication of
neutrino oscillation like the solar neutrino problem or the
atmospheric neutrino anomaly, but the game is not over. Already, some
modifications have been suggested in this picture. One important
point, raised by Bisnovatyi-Kogan \cite{pk|BK}, is that the cross
sections for neutrino interactions are modified in presence of a
magnetic field. Thus, the neutrino spheres themselves will, in
general, be modified, and will in general not remain spherical
surfaces. Following this suggestion, Roulet \cite{pk|roulet} has
performed a careful calculation of the cross section of the process
$\nu_en \to pe$ in presence of a magnetic field. He concludes that for
some ranges of values of the magnetic field and neutrino energy, one
actually needs smaller values of $\Delta m^2$.

\section{Supernova shock and neutrinos}\label{sh}
A supernova is an explosion. A shock wave is formed in the
gravitational collapse of the core of a highly evolved star, which
ejects all surrounding material in space, and we see an explosion. The
problem is that, in computer simulations of these series of events,
the shock was found to be too weak to eject all the surrounding
material. The shock wave stalls after it gets out to a distance of a
few hundred kilometers. If that happens, all material would fall back
and accrete on the dense core already formed, and the result would be
a black hole. Nevertheless, supernovas occur, and therefore it is a
problem to understand what makes the shock strong enough for that to
happen.

We must make a cautionary remark at this point. The simulations, until
very recently, were performed with a one-dimensional model of the
shock wave. Thus, the results may or may not represent the real
situation in three dimensions. Very recently, higher dimensional
simulations have been undertaken, and we should wait for their
results. But in any case, one can be motivated by the one-dimensional
results and try to find out any way of energizing the shock.

Of course, during the gravitational collapse, many neutrinos are
emitted. Some time ago, Bethe and Wilson \cite{sh|BW} argued that
these neutrinos can interact with matter in the form of nucleons or
nuclei in the outer mantle through the reactions
	\begin{eqnarray}
\nu_e + n \to p+e^- \,, && \bar\nu_e + p \to n+ e^+ \,,
\nonumber\\* 
\nu_e + (N,Z) \to (N-1,Z+1)+e^- \,, && \bar\nu_e + (N,Z) \to
(N+1,Z-1)+ e^+ \,,. 
\label{sh.reactions}
	\end{eqnarray}
The mantle is, of course, outside the neutrino-sphere. Thus, the
neutrinos will mostly escape through the mantle. However, a few of
them will indeed intereact as shown above.  This will put extra energy
in the nucleons and nuclei, thereby energizing the shock and
dissociating nuclei ahead of the shock. However, what they found is
that even this is not enough.

After the mechanism of resonant neutrino conversion was proposed to
solve the solar neutrino problem, Fuller, Mayle, Meyer and Wilson
\cite{sh|FMWS} examined whether this can help in the problem of
supernova shock stalling. The point here is that, the $\nu_\mu$'s and
$\nu_\tau$'s, because of smaller cross section with matter, escape
from an inner layer and therefore have larger energy. If they convert
to $\nu_e$ by the resonant conversion mechanism, they will have larger
energy than the original $\nu_e$'s. Thus, if they interact with the
mantle via the reaction of Eq.~\ref{sh.reactions}, they will impart
more energy to the nucleons. Thus, this mechanism will make the shock
revitalization more efficient.

More recently, Akhmedov, Lanza, Petcov and Sciama \cite{sh|ALPS} have
considered another possibility, based on resonant spin-flavor
precession which can take place if the neutrinos have some magnetic
moment. To keep the discussion simple, they assumed that the neutrinos
are Majorana particles, so that no static magnetic moment exists. Only
transition magnetic moments can exist in this case. In the case of two
generations, there is only one independent magnetic moment, the
operator for which connects $\nu_\mu$ with $\bar\nu_e$, and
equivalently $\nu_e$ with $\bar\nu_\mu$.

Bethe and Wilson \cite{sh|BW} already showed that the energy
absorption co-efficients for the reactions in Eq.~\ref{sh.reactions}
is given by 
	\begin{eqnarray} 
K_i (T_\nu) \approx K_0 Y_i T_\nu^2 \,, 
	\end{eqnarray} 
where $K_0$ is a constant, the subscript $i$ stand for either proton
or neutron, $Y_i$ is the relative abundance of $i$, and $T_\nu$ is the
neutrino temperature. Thus, the energy absorption co-efficient in the
case considered by Bethe and Wilson is 
	\begin{eqnarray} 
\dot E \sub{BW} = K_0 \left(  Y_n T_{\nu_e}^2 + Y_p T_{\bar\nu_e}^2
\right) \,,
	\end{eqnarray}
assuming the heating is only
through the free nucleons. 
On the other hand, if resonant spin-flavor precession takes place, the
$\bar\nu_e$'s can come from $\nu_\mu$'s, as indicated above. So, in
that case, one would obtain the energy absorption rate to be
	\begin{eqnarray} 
\dot E \sub{ALPS} = 
K_0 \left( Y_n T_{\nu_e}^2 + Y_p T_{\nu_\mu}^2 \right) \,.
	\end{eqnarray}

Because of larger cross section, the $\bar\nu_e$'s escape from a
sphere further from the center of the proto-neutron star at the core
compared to the $\nu_\mu$'s. Thus, they have a lower temperature,
i.e., $T_{\bar\nu_e}<T_{\nu_\mu}$. Hence the resonant spin-flavor
mechanism must be more efficient in the reheating. Using the values
	\begin{eqnarray}
\langle E_{\nu_e} \rangle &\approx& \phantom{1} 9 \; {\rm MeV},
\nonumber\\* 
\langle E_{\bar\nu_e} \rangle &\approx& 12 \; {\rm MeV},\nonumber\\*
\langle E_{\nu_\mu} \rangle &\approx& 20 \; {\rm MeV},
	\end{eqnarray}
they obtained
	\begin{eqnarray} 
{\dot E \sub{ALPS} \over \dot E \sub{BW}} \approx 2.1 \,,
	\end{eqnarray}
using $Y_p\approx 0.47$ and $Y_n\approx 0.53$. 

A few comments are in order. The mechanism requires that the resonance
takes place outside the neutrino-spheres ($r\sim 50$\,km) and inside the
position of the stalled shock ($r\sim 400$\,km). For small vacuum
mixing angles and for the neutrino energies mentioned above, this
requires~\cite{sh|ALPS} 
	\begin{eqnarray}
10 \; {\rm eV}^2 < \Delta m^2 < 4\times 10^5 \; {\rm eV}^2 \,.
	\end{eqnarray}
Interestingly, the lower end of this range would not conflict with any
cosmological constraints. Moreover, they argue that for such small
$\Delta m^2$, their mechanism is more efficient than the one without
any magnetic moment.

With the range of $\Delta m^2$ given above, and assuming a magnetic
field of the form 
	\begin{eqnarray}
B_\perp (r) = B_0 \left( {r_0 \over r} \right)^k
	\end{eqnarray}
where $r_0$ is the radius of the neutrino-sphere and $B_0$ is the field
at the neutrino-sphere, they can obtain a lower bound for the
transition magnetic moment $\mu$ which ensures that the transition is
adiabatic. For $B_0=5\times 10^{14}$\,G and $k=2$, this gives
	\begin{eqnarray}
\mu \geq (10^{-14} \mbox{ to } 10^{-13} ) \times \mu_B \,.
	\end{eqnarray}
It is not difficult to construct particle physics models which predict
neutrino magnetic moments in this range.

\section{Neutrino astronomy}\label{as}
Since the birth of astronomy, we have detected light from distant
objects to find out the nature of these objects. In the twentieth
century, the detection was extended to other parts of the
electromagnetic spectrum, so that now we have x-ray, infra-red and
radio astronomy. Within the last quarter of a century, the detection
went beyond the electromagnetic spectrum by beginning to detect
neutrinos. This endeavor started with the detection of solar
neutrinos in the 1970's, and those experiments are still going on. In
1987, neutrinos from a supernova was also detected, and neutrino
astronomy has now come of age.

Neutrino astronomy has its advantages and disadvantages over photon
astronomy. The main disadvantage is that, since neutrinos have much
smaller cross section with the detector as compared with the photons,
one needs large detectors. But the advantages are many. Neutrinos
suffer hardly from any distraction during their journey. They arrive
directly in line from the source. They can bring astrophysical
information from cores of various object (like the sun) which photons
cannot.

Since my talk excludes solar neutrinos, I will not discuss various
operating as well as upcoming solar neutrino detectors. I will discuss
another class of detectors which were inspired by the success of the
solar neutrino detectors as well as the observation of neutrino pulse
from SN1987A. These are detectors for Ultra High-Energy (UHE)
neutrinos.

There are two questions about the UHE neutrino telescopes: (1) what
kind of new phenomenon will be observable by them; and (2) what kind
of event rates can one expect.

As for the first question \cite{as|gaisser}, one might expect to
detect the diffuse 
neutrino emission from our galaxy. There are also interesting
extragalactic sources, and we list a few:
	\begin{itemize}

\item Active galactic nuclei (AGNs): These are regions of new star
	formation at the center of galaxies. Protons and electrons are
	accelerated to high energy by shock waves. The charged
	particles remain trapped by the diffuse magnetic field. But
	there are reactions of the type $p+\gamma\to n+\pi^+$, and
	neutrons escape to form cosmic rays. Neutrinos are created
	from charged pion decays, and their energies will be
	comparable to those of the cosmic rays.
	
\item Gamma ray bursters (GRBs): These are sources of huge gamma ray
	bursts, which are suspected to occur due to merger of neutron
	stars. 
\item Topological defects (TDs): If there are topological defects like
	cosmic strings, we expect neutrino fluxes from them.
	\end{itemize}
There may also be other unexpected sources. But let us now turn our
attention to the second question.

The answer to the second question depends on the interaction of
neutrinos with nucleons and electrons which constitute detector
material. Calculation of the cross section with nucleons require
knowledge of nucleon structure functions. The structure functions are
functions of two variables. One of them is usually taken to be
$Q^2=-q^2$, where $q^\mu$ is the 4-momentum exchanged between the
neutrino and the nucleon. The other is 
the Bjorken variables $x$, which, in the rest frame of the interacting
nucleon, is given by
	\begin{eqnarray}
x = {Q^2 \over 2 m_N E} \,,
	\end{eqnarray}
$E$ being the energy carried off by the
intermediate vector boson. For UHE neutrinos, $E\approx E_\nu$, the
energy of the incoming neutrinos. Thus, we want structure functions at
very small $x$. For example, if one wants to consider $E_\nu\sim
10^9$\,GeV, one needs structure functions at $x\sim 10^{-6}$.

So far, no experiment has measured structure functions to such low
values of $x$. The lowest values of $x$ have been probed by the
$ep$-collider HERA, which can go as low as about $10^{-4}$, and these
HERA results have been available only very recently. In
order to find cross sections for smaller values than this, one needs
to extrapolate these results.

Gandhi, Quigg, Reno and Sarcevic \cite{as|GQRS} have performed
extensive analysis of the known regime of structure functions and
extrapolated them to smaller $x$. With these extrapolations, they
calculated neutrino-nucleon cross sections, and found that their
results are substantially higher than the ones calculated with earlier
extrapolations of structure functions \cite{as|EHLQ}. The reason for
the difference is twofold. First, they used the the structure
functions derived by the CTEQ collaboration \cite{as|CTEQ} from the
HERA results which were not known earlier. Second, they use a mixture
of various extrapolation techniques to make the extrapolation more
reliable for small $x$. Their results now form the standard framework
in which the cross sections of UHE neutrino detectors are calculated.
With their results, we now present the event rate expected from
various sources mentioned earlier. This appears in
Table~\ref{t:eventrates}.

\begin{table}
\caption[]{\sf Upward $\mu^++\mu^-$ event rates per year for all nadir
angles for a detector with effective area $0.1$\,km$^2$, with two
different values of the threshold energy. From
Ref.~\cite{as|raj}.\label{t:eventrates}} 
\begin{center}
\begin{tabular}{lcr@{---}l|r@{---}l}
\hline \hline
\multicolumn{1}{c}{Flux} & Ref. & \multicolumn{4}{c}{Threshold
$\mu$-energy} \\  \cline{3-6}
&& \multicolumn{2}{c|}{1\,TeV} & \multicolumn{2}{c}{10\,TeV} \\  \hline 
AGN & \cite{as|M95} & 31 & 33 & 6 & 7\\
AGN ($p\gamma$) & \cite{as|P96} & 54 & 56 & 29 & 37 \\
AGN ($p\gamma$ \& $pp$) & \cite{as|P96} & 2130 & 2258 & 433 & 479 \\
GRB & \cite{as|GRB} & 12 & 13 & 5 & 6 \\ 
TD & \cite{as|TD} &\multicolumn{2}{c|}{0.007}  \\ 
\hline \hline
\end{tabular}
\end{center}
\end{table}
In viewing this table, one has to remember the following. The fluxes
of neutrinos from various kinds of sources described above is not
well-known. There are several calculations of neutrino fluxes from
AGNs, for example. We have therefore presented the expected event
rates corresponding to these different results. Also, the first
calculations for AGNs considered neutrinos created from the $p\gamma$
reactions. A recent calculation also put in neutrinos created from
$pp$ reactions. This increases the expected fluxes fantastically, as
can be seen from Table~\ref{t:eventrates}.

Because these numbers are accessible to experiments, a few experiments
are planned. These are all under-water or under-ice detectors. The
AMANDA detector at the south pole has been completed recently. The
others, which are at various stages of developments, are (1) 
BAIKAL neutrino telescope, at a depth of 1\,km in Lake Baikal in
Siberia; (2) NESTOR, at a depth of  3.5\,km in the Mediterranean near
Pylos, Greece. Another one, DUMAND, at a depth of 4.7\,km in the ocean
30\,km off the island of Hawaii, has been abandoned midway.

\section{Neutrinos in cosmology}\label{co}
The importance of neutrinos in cosmology derives from the fact that
they are the most dominant particles in the universe, apart from
photons. It was believed for a while that they could be the dark
matter of the universe, for which various indications exist at various
scales. These indications will not be reviewed here.

At first, it was believed the neutrinos can constitute all of the dark
matter in the universe. Later it was realized that in such a universe
filled with light neutrinos, it is difficult to form structures. An
alternative, cold dark matter scenario was favored then. But neutrinos
staged a comeback with the publication of the COBE data, which showed
that not enough structure can be made with CDM at large scales. Now,
it is believed that hot dark matter constitutes about 20--30\% of the
universe, and of course neutrinos are the prime candidates for hot
dark matter.  This, by now, is part of the folklore, and so I will not
get into details. I will rather talk about something recent, as
promised.

\begin{figure}
\centerline{ 
\epsfysize=0.7\textheight
\epsfbox{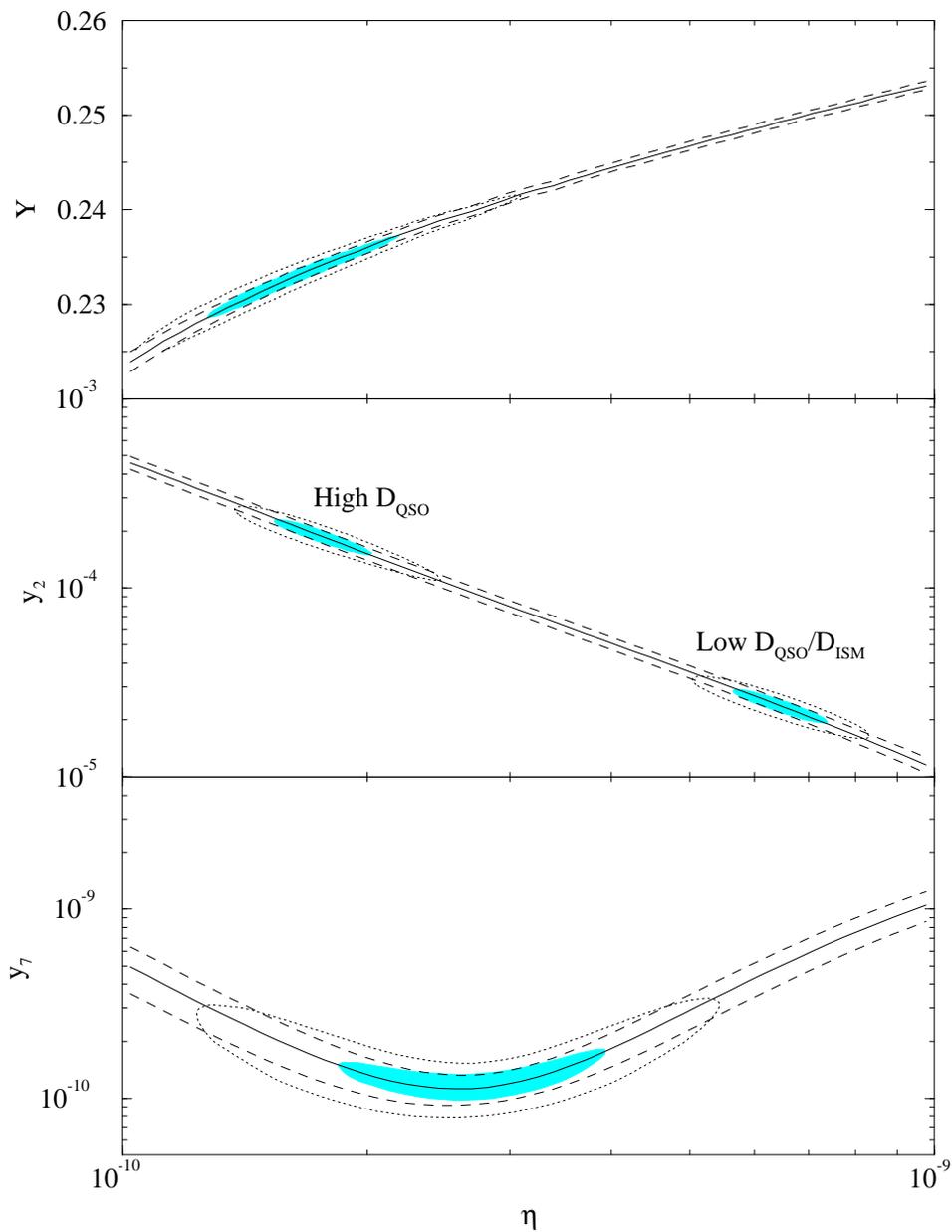}
}

\caption[]{\sf The dependence of various abundances on the parameter
$\eta$ are shown with dotted lines. The top panel shows the abundance
of Helium, the middle panel of deuterium, and the lower panel of 
Lithium. The observations are marked on this plot. The new data
appears in the lower right end of the middle panel. From 
Ref.~\cite{co|2nu}.}\label{SHBLfig} 

\end{figure}
There are some recent measurements \cite{co|obs} of deuterium
abundance in the universe which imply a much 
lower value of the quantity than was believed before. If this value of
the deuterium 
abundance is believed, this implies a larger value of the parameter
$\eta$ which stand for the baryon to photon number in the universe, as
seen from Fig.~\ref{SHBLfig}.

Steigman, Hata, Bludman and Langacker \cite{co|2nu} have explored the
possible implications of this observation. If we take the value of
$\eta$ dictated by this observation, it would imply that the
primordial Helium abundance is much larger than what was believed so
far. This higher value would be inconsistent with the observations on
primordial Helium abundance.

However, the plots in Fig.~\ref{SHBLfig} assume three massless
neutrino flavors contributed to the energy density of the universe at
the time of Helium synthesis. If, instead, we assume the number of
flavors to be two, a better agreement is obtained.

Of course, we know that there are three kinds of neutrinos, $\nu_e$,
$\nu_\mu$ and $\nu_\tau$. The direct measurements of the masses of
these particles indicate that the $\nu_e$ mass must be smaller than a
few eV, and the $\nu_\mu$ mass should be smaller than about
250\,keV. Since the Helium synthesis occurred when the temperature was
about an MeV, both these neutrinos must have been effectively massless
at that time. However, the experimental upper limit of $\nu_\tau$ is
23\,MeV. If the mass is really close to that upper limit, $\nu_\tau$
would not count as an effectively massless species. And then the number
of effectively massless neutrino species would be two.

Thus, if we take this piece of data seriously, one implication is that
the $\nu_\tau$ mass should be larger than an MeV. One can of course
argue about how reliable is the data. Or how reliable are the data on
primordial Helium and Lithium abundance. I am not qualified to make a
comment on this issue.

\paragraph*{Acknowledgments~:} I thank Raj Gandhi for providing me
with unpublished results which have been presented in
Table~\ref{t:eventrates}. I also thank Manuel Drees for asking me some
questions during and after the talk, and Kamales Kar for his comments
on an earlier draft, all of which have helped me understand the
subject of \S\,\ref{sh} better. I thank S. Bhowmik-Duari,
D. Bandopadhyay and S. Pal for various discussions on the topics in
\S\,\ref{sh} and \S\,\ref{co}.


\begin{thebibliography}{[WW]}
\bibitem{pk|DNPnote} The results of  Ref.~\cite{pk|DNP} are presented
in a convention in which $e$ is the charge of the electron, i.e., is
negative. This is the convention used by Ref.~\cite{pk|KS} also. Here,
we use a convention in which $e>0$, which is more standard, and has
been used in some of the papers in Ref.~\cite{pk|reso}.
\end{thebibliography}
\end{document}